\documentclass[11pt,graphicx,amsmath]{article}
\usepackage{amsmath}
\usepackage{graphicx}
\usepackage{bm}
\usepackage[dvips]{color}
\usepackage{amssymb}
\usepackage{amsfonts}
\usepackage{comment}
\usepackage{cite}
\usepackage{todonotes}
\usepackage{caption}
\usepackage{subcaption}

\def\be{\begin{equation}}
\def\ee{\end{equation}}
\def\ba{\begin{eqnarray}}
\def\ea{\end{eqnarray}}
\def\nn{\nonumber}

\def\bl#1\el{\begin{align}#1\end{align}}

\title{ Maxwell field with gauge fixing term
in the radiation- and matter-dominant stages:
exact solution and stress tensor }
\author{\small
           Xuan Ye \thanks{yyyyy@mail.ustc.edu.cn}  \,
                 and Yang Zhang  \thanks{yzh@ustc.edu.cn}
            \\
 \small  Department of  Astronomy,
         CAS Key Laboratory for Researches in Galaxies and Cosmology, \\
 \small  School of Astronomy and Space Sciences,
         University of Science and Technology of China, \\
         Hefei, Anhui, 230026, China \\
 }

\date{}

\def\be{\begin{equation}}
\def\ee{\end{equation}}
\def\ba{\begin{eqnarray}}
\def\ea{\end{eqnarray}}
\def\nn{\nonumber}
\def\bl#1\el{\begin{align}#1\end{align}}

\large

\begin{document}

\maketitle
\allowdisplaybreaks

\begin{abstract}

\large

We study the Maxwell field with a general gauge fixing (GF) term
in the radiation-dominant (RD)  and matter-dominant (MD) stages
of expanding Universe,
as a continuation to the previous work in the de Sitter space.
We derive the exact solutions,
perform  the covariant canonical quantization
and obtain the stress tensor in the Gupta-Bleuler (GB) physical states,
which is independent of the GF constant
and is also invariant under the quantum residual gauge transformation.
The transverse stress tensor is similar in all flat Robertson-Walker spacetimes,
and its vacuum part is $\propto k^4$
 and becomes zero after the 0th-order adiabatic regularization.
The longitudinal-temporal stress tensor,  in both the RD and MD stages,
is zero due to a cancelation between the longitudinal and temporal parts
in the GB states,
and so is the particle part of the GF stress tensor.
The vacuum  GF stress tensor,  in the RD stage,
contains $k^4,k^2$ divergences
and becomes zero by the 2nd-order  regularization,
however, in the MD stage, contains $k^4, k^2, k^0$ divergences
and becomes zero by the 4th-order regularization.
So, the order of adequate regularization
depends not only upon the type of fields, but also upon the background spacetimes.
In summary,  in both the RD and MD stages, as in the de Sitter space,
the total regularized vacuum stress tensor is zero,
 independent of the GF constant,
only the transverse photon part remains,
there is no trace anomaly,
and the vanishing GF stress tensor
can not be a candidate for the dark energy.

\end{abstract}

\large

\section{Introduction}

For  the covariant canonical quantization of the Maxwell field,
the GF term $\frac{1}{\zeta}(\nabla_{\mu}A^{\mu})^2$
is commonly introduced into the Lagrangian density
\cite{ItzyksonZuber,AdlerLiebermanNg1977,BrownCassidy1977}.
This GF term not only mixes up
the longitudinal and temporal components  $A$ and $A_0$ in the field equations,
but also may cause unwanted consequences in curved spacetimes,
such as the appearance of the GF stress tensor,
which seemingly depends on the GF constant
and might be a possible candidate for the cosmic dark energy
as suggested in Refs.\cite{JimenezMaroto2009,JimenezMaroto2010}.
From the point of view of a field theory   \cite{Piguet1985},
the GF term is purely auxiliary and introduced
for the purpose of covariant quantization,
and no measurable effect can be associated to it.
Often a ghost field is additionally introduced
\cite{AdlerLiebermanNg1977,BrownCassidy1977},
in particular, Ref.\cite{AdlerLiebermanNg1977} used
the ghost field to cancel the GF stress tensor
in the Feynman gauge $(\zeta=1)$.
Refs.\cite{JimenezMaroto2009,JimenezMaroto2010}
took the GF term as  the dark energy
but  did not address the issue of its vacuum UV divergences.
As we shall show in this paper,
when the regularization is applied to remove the UV divergences,
the vacuum GF stress tensor becomes zero,
having null physical effects    \cite{Piguet1985}.
Ref. \cite{Vollick2012} considered the Maxwell field in the Lorenz gauge
(i.e, without the GF term) and set the vacuum stress tensor to zero
by the normal ordering which is a procedure inappropriate for quantum fields
in curved spacetimes as pointed in Refs.\cite{FeynmanHibbs1965,DeWitt1975}.
Ref. \cite{Glavan2212.13982} studied the Maxwell field with the gauge fixing term
in D-dimensional RW spacetimes,
and used the operator ordering which is actually equivalent to the normal ordering.
Ref.\cite{NiardiEspositoTramontano2021}
studied the Maxwell field with the gauge fixing term
using the point-splitting regularization based on the Schwinger-DeWitt expansion,
and isolated the UV divergences of the stress tensor
in the Ricci-flat and the maximally symmetric spaces.
Ref. \cite{AllenJacobson1986} studied the Proca field and
Ref. \cite{FrobHiguchi2014}  studied the Stueckelberg field in the de Sitter space,
and  calculated the two-point function,
but the stress tensors were not addressed.
Ref.\cite{ChuKoyama2017} calculated the trace
of the stress tensor of the Stueckelberg field in a RW spacetime,
and applied the 4th-order adiabatic regularization,
but did not give all the components of the stress tensor.
The Maxwell field with the gauge fixing term is a gauge field,
and is much more complicated
than the Stueckelberg field which is not a gauge field.
For the Maxwell field, the solutions of $A$ and $A_0$ are more involved,
and these are unphysical degrees of freedom
and need to be removed in physical quantities, such as the stress tensor.
Besides, the residual gauge transformation
is further restricted by the canonical quantization.
For the Stueckelberg field,
all four components $A_\mu$ are dynamical degrees of freedom,
and the solutions of $A$ and $A_0$
can be inverted,  by the massiveness,
 from the canonical momenta
by algebraic combinations and differentiations.
In our previous  study \cite{ZhangYePRD2022}
of the Maxwell field with a general GF term
in the de Sitter space and in the  Minkowski spacetime,
we have derived the solutions of $A$ and $A_0$,
implemented the covariant canonical quantization,
and obtained the total stress tensor in the GB physical states.
The longitudinal and temporal (LT) stress tensor
and the photon part of the GF stress tensor are vanishing
in the GB states.
Only the transverse stress tensor and the vacuum GF stress tensor remain,
and are amazingly independent of the GF constant $\zeta$,
and invariant under the quantum residual gauge transformation
as a subset of the residual gauge transformation.
But the vacuum transverse  and GF stress tensors contain UV divergences.
By the adiabatic regularization,
we found that the vacuum transverse stress tensor becomes zero
by the 0th  order regularization,
and the vacuum GF stress tensor is zero by the 2nd order regularization.
As a result, the total regularized vacuum stress tensor is zero,
only the photon part of the transverse stress tensor remains,
and all the physics predicted by the Maxwell field
with the GF term will be the  same as that without the GF term.
In particular,
given the vanishing vacuum GF stress tensor,
it can not contribute to  the dark energy,  nor to the cosmological constant,
and one needs no introduction of a ghost field,
which itself would cause additional unphysical consequences.

In this paper, we  study the Maxwell field
with the GF term in another two flat Robertson-Walker (fRW) spacetimes:
the RD and  MD stages of the expanding Universe,
both stages being of importance in cosmology.
The transverse components are simple
in a  fRW spacetime,
but the longitudinal and temporal components are nontrivial
as they are mixed up.
Once the complete set of solutions are derived,
the covariant canonical quantization,
the calculation of the stress tensor,
and the adiabatic regularization on the vacuum stress tensor will be carried out.
We shall show that, in the RD and MD stages,
the regularized vacuum stress tensor of the Maxwell field is zero,
and there exists no trace anomaly for the Maxwell field,
thus disproving the claim in literature
that the Maxwell field had a trace anomaly
\cite{BrownCassidy1977,DowkerCritchley1977,ChuKoyama2017}.
Our calculation confirms that the trace anomaly will never occur
when one works  directly with a massless field
 \cite{ZhangYePRD2022, ZhangYeWang2020, ZhangWangYe2020, YeZhangWang2022}. 
The Maxwell field in these two stages
provides two other  specific examples
that  the trace anomaly will not arise in curved spacetimes.
Moreover,
our calculation will demonstrate that the GF stress tensor in the MD stage
requires the 4th-order regularization,
instead of the 2nd-order in the RD stage and the de Sitter space,
and thus providing an important example of the general statement
that an appropriate regularization scheme, for a given field,
generally depends on the background spacetimes.
The treatments in the RD and MD stages in this paper
are quite similar,
and also analogous to the de Sitter space \cite{ZhangYePRD2022}.
So we shall concisely report the results for the RD stage in the context,
and list the main results for the MD stage in the Appendix.

The paper is organized as follows.
Sect.2, we derive the solutions of the Maxwell field
with the gauge fixing term in the RD stage.
Sect.3. we implement the covariant canonical quantization.
Sect.4, we calculate all the three parts of the stress tensor.
Sect.5, we perform the adiabatic regularization on the vacuum stress tensor.
Sect.6  gives the  conclusion  and  discussion.
Appendix A lists the results for the MD stage.

\section{The solutions in RD stage}

We firstly give the setup and notations as in \cite{ZhangYePRD2022}.
The metric of a fRW spacetime is
$ds^2 =a(\tau)^2[-d\tau^2+\delta_{ij}(d\vec{x})^2]
=a(\tau)^2\eta_{\alpha\beta}dx^\alpha dx^\beta $,
where  $\tau$ is the conformal time.
The Lagrangian density of the
Maxwell field with the GF  term in a curved spacetime is
${\cal L}  = \sqrt{-g} \big( -\frac14 g^{\mu\rho}g^{\nu\sigma}
    F_{\mu\nu} F_{\rho\sigma}   -\frac{1}{2\zeta}
     (\nabla^\mu A_\mu )^2  \big)$,
where $F_{\mu\nu}= A_{\mu, \, \nu}- A_{\nu,\, \mu}$,
and $\zeta$ is the gauge fixing constant.
The field equation  is
\bl
F^{\mu\nu}_{~~;\nu}+\frac{1}{\zeta}(\nabla^\nu A_\nu )^{;\mu} = 0.
\label{feidequationcov}
\el
The Lorenz  condition will not be imposed on the field operators
in this paper.
The $i$ component is decomposed into
the  transverse  and longitudinal parts $A_i=\partial_{i}A+B_i$,
where $B_i$ satisfies  $\partial_{i}B_i=0$.
The canonical momentum is given by
\bl
\pi^{\mu}_A=\frac{\partial{\cal L}}{\partial(\partial_0 A_{\mu})}
=\eta^{\mu\sigma}(\partial_0A_\sigma-\partial_\sigma A_0)
 -\frac{1}{\zeta} \eta^{0\mu}(\eta^{\alpha\beta}\partial_{\alpha}A_{\beta}
 -DA_0),
\el
with $D \equiv 2a(\tau)'/a(\tau)$,
and $a(\tau)' \equiv d a(\tau)/d\tau$,
the prime  denoting the derivative with respect to the conformal time.
The $i$ component is also decomposed into
$\pi^i_A =w^i +\partial^i \pi_A$  with $w^i=\partial_0 B_i$ and
\bl
\pi_A =\partial_0A-A_0 ,\label{definationpirealspace}
\el
and the $0$ component  is contributed by the GF term
\bl
\pi^0_A =\frac{1}{\zeta}a^2A^{\nu}_{~;\nu}
   =-\frac1\zeta\big((\partial_0+D)A_0-\partial_i^2A\big).
\label{definationpi0realspace}
\el
In the following we shall work with the $k$ space,
and use $B_i$, $A$,  $A_0$, $\pi_A$, $\pi_A^0$
to represent their Fourier $k$ modes whenever no confusion arises.
Then, the $k$ mode of \eqref{definationpi0realspace} is
\bl
\pi^0_A =
-\frac1\zeta\big((\partial_0+D)A_0+k^2A\big).
\label{definationpi0kspace}
\el
So, Eq.\eqref{feidequationcov}
is decomposed into the following three equations in the $k$ space
\bl
 \partial_0^2B_i+k^2 B_i & = 0,
\label{equationtranse}
\\
  -   \partial_0^2   A
     -\frac{1}{\zeta} k^2  A
 + \Big(1- \frac{1}{\zeta} \Big)\partial_0  A_0
    - \frac{1}{\zeta} D A_0 &  = 0,
\label{equationAA02}
    \\
 - \frac{1}{\zeta}\partial_0^2  A_0 -k^2 A_0
  +\frac{1}{\zeta} ( D^2   -D' ) A_0
 +k^2 \Big[ (1- \frac{1}{\zeta} )\partial_0  A
 + \frac{1}{\zeta}  D A   \Big] & =0  ,
\label{equationAA01}
\el
 with $D' \equiv d D(\tau)/d\tau$.
The transverse $B_i$ are independent of $\zeta$ and separate from $A$ and $A_0$,
and have the positive frequency solutions
\bl
B_{i} \propto f^{(\sigma)}_k(\tau)
   =\frac{1}{\sqrt{2k}}e^{-ik\tau},\label{solutionoftranseverse}
\el
where  $f_{k}^{(\sigma)}$
are the same for two transverse polarizations $\sigma=1,2$,
and  hold for a general fRW spacetime.
Eqs. \eqref{equationAA02} and \eqref{equationAA01}
are two basic second-order differential equations,
in which $A_0$ and $A$ are mixed up for a general $\zeta$.
Combination of \eqref{equationAA02} and  \eqref{equationAA01}
leads to two fourth-order differential equations
\bl
& \Big[ \Big(1- \frac{1}{\zeta} \Big)\partial_0
- \frac{1}{\zeta} D  \Big]
\Big( \frac{[ (\zeta - 1) \partial_0^3
  +   D \partial_0^2
 + k^2 (\zeta - 1) ( 2 - \zeta )\partial_0
   +  (2 - \zeta)  k^2 D ]  A}
   {(\zeta - 2)  D^2 -(\zeta - 1) D' - (\zeta - 1)^2 k^2} \Big)
   - \Big(  \partial_0^2   + \frac{1}{\zeta} k^2 \Big)  A  = 0 , \label{Aeom}
\el
and
\bl
&  \Big[ \Big(1- \frac{1}{\zeta} \Big)\partial_0 + \frac{1}{\zeta} D \Big]
     \Big( \Big[ (\zeta -1) D' - D^2  -\frac{(\zeta -1)^2}{\zeta}  k^2 \Big]^{-1}
     \Big[ (\zeta -1) \partial_0^3 A_0 - D  \partial_0^2  A_0
  \nn \\
&  + (\zeta -1) \Big( D' - D^2 - \frac{1-2\zeta }{\zeta} k^2 \Big) A_0'
  + (\zeta -1)  ( D'' - 2 D D' ) A_0
\nn
\\
&  +   D ( D^2  -D' ) A_0
    + \frac{1-2\zeta }{\zeta} D k^2  A_0 \Big]\Big)
     -k^2 A_0 - \frac{1}{\zeta}\partial_0^2  A_0
  +\frac{1}{\zeta}  ( D^2  -D'  ) A_0=0 ,\label{A0eom}
\el
with  $D(\tau)''\equiv  d^2 D(\tau) / d\tau^2$.
Eqs.\eqref{Aeom} \eqref{A0eom} are separated for
  $A$ and $A_0$, and valid for $\zeta \ne 1$.
The preceding setup holds for a fRW spacetime  \cite{ZhangYePRD2022}.

Now we consider the RD stage,   the scale factor is
\bl
a(\tau)=a_{r} \tau, \label{salcefactor}
\el
where $a_{r}$ is a constant.
Dropping an overall factor $\propto(1-\zeta)^2$,
 Eqs. \eqref{Aeom} and  \eqref{A0eom} for the RD stage become
\bl
&[(\zeta -1)^2 k^2 \tau ^2 - 2(3 \zeta -5)] \tau ^4 A^{(4)}(\tau )
   - 4 (3 \zeta   -5)\tau^3  A^{(3)}(\tau )\nn
\\
&~~~~ + 2 [(\zeta -1)^2 k^4 \tau ^4-  k^2 \tau ^2
      (\zeta  (3 \zeta +2)-9)+6 (3\zeta -5)] \tau^2 A''(\tau )
   + 4 (\zeta -2) k^2 \tau ^2  (3 \zeta   -5 )\tau A'(\tau )
\nn
\\
&~~~~ + k^2 \tau ^2 [ (\zeta -1)^2 k^4 \tau ^4
      +2 k^2   \tau ^2 (\zeta  (\zeta -7)+8) + 4(3\zeta -5) ]  A(\tau )=0,
  \label{4th0rdersolutionA}
\el
and
\bl
&[(\zeta -1)^2 k^2 \tau ^2+2 \zeta  (\zeta +1)]\tau ^4 A_0^{(4)}(\tau )+4 \zeta    (\zeta +1)\tau ^3 A_0^{(3)}(\tau )\nn
\\
&~~~ +2  [(\zeta -1)^2 k^4 \tau ^4+  k^2\tau ^2 (\zeta  (10- \zeta)-5)
     -6 \zeta (\zeta +1)]\tau ^2 A_0''(\tau )
\nn
\\
&~~~ +4    [ k^2 \tau ^2 (\zeta  (8\zeta -11)+5) + 6 \zeta (\zeta +1)] \tau A_0'(\tau )
\nn
\\
&~~~ + [ (\zeta -1)^2 k^6 \tau ^6 + 2 k^4 \tau ^4 (2 \zeta ^2 + \zeta -1)
     -4 k^2\tau ^2 (\zeta  (8\zeta  -11)+5) - 24 \zeta  (\zeta +1) ] A_0(\tau )
    =0,\label{4th0rdersolutionA0}
\el
where $A^{(4)}(\tau )\equiv \partial^4A/\partial\tau^4$,
$A^{(3)}(\tau )\equiv \partial^3A/\partial\tau^3$, etc.
Eqs. \eqref{4th0rdersolutionA} and  \eqref{4th0rdersolutionA0}
are linear equations with the coefficients being power of $\tau$,
and hold for a general $\zeta$,
and the positive frequency solutions  are given by
\bl
A&=c\frac{1}{a(\tau)}\frac{i}{k}\frac{1 }{\sqrt{2k}}e^{-ik\tau}
+\alpha\frac1k\frac{3\zeta +15+6ik\tau(\zeta +5)
-6k^2\tau^2(\zeta +1)
-4ik^3\tau^3(\zeta -1)}{24  k\tau}
 \frac{1}{\sqrt{2k}} e^{-i k\tau},\label{Arsolution4th}
\\
A_{0}&=c\frac{1}{a(\tau)}\frac{1 }{\sqrt{2k}} \Big(1-\frac{i}{k\tau}\Big)
e^{-ik\tau}+\alpha\frac{ 4 k^4 \tau ^4 (1-\zeta )
- (2 i k^3 \tau ^3+3i k \tau +3) (\zeta +5)}{24  k^2\tau^2 }
\frac{1}{\sqrt{2k}} e^{-i k \tau },\label{A0rsolution4th}
\el
where the coefficients $c,~\alpha$ are dimensionless complex constants.
At the classical level, $(c, \alpha)$ are arbitrary,
however, they will be
further constrained at the quantum level at Sect.\ref{quantization}.
We have chosen the same set of coefficients  $(c, \alpha)$ for $A$ and  $A_0$,
so that \eqref{Arsolution4th} and \eqref{A0rsolution4th}  satisfy
the basic second-order equations
\eqref{equationAA02} and \eqref{equationAA01}.
Substituting  \eqref{Arsolution4th} and \eqref{A0rsolution4th} into
the definitions \eqref{definationpirealspace} and \eqref{definationpi0kspace}
gives the canonical momenta
\bl
\pi_A &=\alpha\frac{a(\tau)}{a_{r}}ik\frac{1 }{\sqrt{2k}}
    \Big(1-\frac{i}{k\tau}\Big) e^{-ik\tau},\label{pir}
\\
\pi^{0}_A &=\alpha\frac{a(\tau)}{a_{r}} k^2 \frac{1}{\sqrt{2k}}e^{-ik\tau}  ,
 \label{pir0}
\el
which are contributed only by the $\alpha$ part of $A$ and $A_{0}$,
and are independent of $\zeta$.

The solutions \eqref{Arsolution4th}, \eqref{A0rsolution4th},
\eqref{pir}, \eqref{pir0} can be also derived in another way in the following.
By combinations of the basic equations
\eqref{equationAA02} and \eqref{equationAA01},
we get the following equations of the canonical momenta
\bl
 (\partial_0^2  - D \partial_0  + k^2 ) \pi_A   =0 ,\label{equationofpiA}
 \\
 ( \partial_0^2    - D \partial_0  -  D'
       +k^2 ) \pi^0_A      =0 .\label{equationofpiA0}
\el
By resealing $\pi_A=a\bar{\pi}_A$ and  $\pi^0_A=a\bar{\pi}^0_A$,
Eqs.\eqref{equationofpiA}  and \eqref{equationofpiA0} become
\bl
\partial_0^2\bar{\pi}_A
+  (k^2-\frac{2}{\tau ^2})\bar{\pi}_A=0\label{equationofpiAres},
\\
\partial_0^2\bar{\pi}^{0}_A
+  k^2\bar{\pi}^{0}_A=0\label{equationofpiA0res},
\el
the solutions are  found as
\bl
\bar{\pi}_A
&=\alpha\frac{1}{a_{r}} i k
\frac{1 }{\sqrt{2k}}(1-\frac{i}{k\tau})e^{-ik\tau},\label{pirpirbar}
\\
\bar{\pi}^{0}_A
&=\alpha\frac{1}{a_{r}} k^2
 \frac{1}{\sqrt{2k}}e^{-ik\tau},\label{pi0rpi0rbar}
\el
which confirm  \eqref{pir} and  \eqref{pir0}.
By combinations of
the definitions \eqref{definationpirealspace}  and \eqref{definationpi0kspace},
we get the following  inhomogeneous equations
\bl
 \partial_0^2 A  + D \partial_0 A   +k^2 A
  &  = (\partial_0 + D)  \pi_A  - \zeta \pi^0_A  , \label{AEu2}
  \\
 \partial_0^2  A_0 +   D A_0'  + D' A_0   + k^2   A_0
     &  = -( k^2  \pi_A + \zeta \partial_0   \pi^0_A ) . \label{A0epi}
\el
By the standard formula and using the known $\pi_A$ and $\pi_A^0$,
the solutions of \eqref{AEu2} and \eqref{A0epi} in the RD stage
are the same as  \eqref{Arsolution4th} and \eqref{A0rsolution4th},
where the $c$ parts are identified as the general homogeneous solutions
and the $\alpha$ parts are the inhomogeneous solutions.

The Maxwell field with  the GF term
is invariant under the residual gauge transformation
$A_\mu \rightarrow A_\mu' \equiv  A_\mu +\theta_{, \, \mu}$,
where  $\theta$ satisfies the equation
$\Box \theta \equiv \nabla^\nu\nabla_\nu \theta=0$.
In the RD stage the $k$-mode solution is
\bl
\theta_k= C \frac{1}{a(\tau)} \frac{i}{k}
           \frac{1}{\sqrt{2k}} e^{-ik\tau},\label{soltheta}
\el
with $C$ being an arbitrary complex constant.
The transverse components  and
the canonical momenta are invariant,
$B_i \rightarrow   B_i$,
$\pi_A  \rightarrow  \pi_A $,
$ \pi^{0}_A  \rightarrow  \pi^{0}_A$.
The longitudinal and temporal $k$-modes transform as
\bl
A_{k} &  \rightarrow A_{k}+\theta_{k}=
A_k+ C \frac{1}{a(\tau)} \frac{i}{k}
           \frac{1}{\sqrt{2k}} e^{-ik\tau} ,
\label{Atrsfrad}
  \\
A_{0\, k} & \rightarrow A_{0\,k}+\theta_{k,\, 0}
= A_{0\,k}+  C \frac{1}{a(\tau)}
  \frac{1}{\sqrt{2k}} \Big(1-\frac{i}{k\tau}\Big) e^{-ik \tau} .
\label{A0trsfrad}
\el
Comparing  \eqref{Atrsfrad}, \eqref{A0trsfrad}
 with \eqref{Arsolution4th}, \eqref{A0rsolution4th},
the residual gauge transformation
amounts to a change of the coefficients of
the homogeneous parts of $A_k$ and $A_{0k}$
as the following
\bl
c\rightarrow c'=c+C.
\el
As we shall show, the parameter $C$ of residual gauge transformation
 will be further restricted by the covariant canonical quantization.

\section{The covariant canonical quantization
in  RD stage}\label{quantization}

With all the $k$-modes in the RD stage available,
we shall use the conventional method of covariant canonical quantization,
in the same way as in de Sitter space \cite{ZhangYePRD2022}.
(In the b-field formalism
 of covariant quantization for gauge fields  \cite{Ojima1990},
one may extend the space of the field variables
and introduce an auxiliary b-field.
Here, for simplicity, we adopt the conventional method for the Maxwell field.)
The field and canonical momenta operators are written as
\bl
B_i   ({\bf x},\tau)
& =  \int\frac{d^3k}{(2\pi)^{3/2}}
  \sum_{\sigma=1}^2 \epsilon^{\sigma}_{i}(k)
  \left[ a^{( \sigma)} _{\bf k}  f_k^{(\sigma)}(\tau)   e^{i\bf{k} \cdot\bf{x}}
   +a^{( \sigma)\dagger}_{\bf k} f_k^{(\sigma)*}(\tau)
   e^{-i\bf{k} \cdot\bf{x}}\right],\label{expBi}
\\
w^i (\tau, {\bf x})
&  =  \int\frac{d^3k}{(2\pi)^{3/2}}
  \sum_{\sigma=1}^2 \epsilon^{\sigma}_{i}(k)
  \left[ a^{( \sigma)} _{\bf k}  f_k^{(\sigma)'}(\tau) e^{i\bf{k} \cdot\bf{x}}
   +a^{( \sigma)\dagger}_{\bf k} f_k^{(\sigma)*'}(\tau)
   e^{-i\bf{k} \cdot\bf{x}}\right],
   \label{wjexp}
\el
\bl
A &  =  \int\frac{d^3k}{(2\pi)^{3/2}}
  \Big[ (a^{(0)}_{\bf{k}}A_{1k}+a^{(3)}_{\bf{k}}A_{2k})
   e^{i\bf{k}\cdot \bf{x}}+h.c.\Big] ,\label{N26}
   \\
\pi_A  & = \int\frac{d^3k}{(2\pi)^{\frac32}}
  \Big( \big( a_{\bf{k}}^{(0)} \pi_{A\, 1k}
  +a_{\bf{k}}^{(3)}\pi_{A\, 2k} \big)
         e^{i\bf{k}\cdot\bf{x}} + h.c.\Big) \label{N33},
         \\
A_0 &  =  \int\frac{d^3k}{(2\pi)^{3/2}}
 \Big[ (a^{(0)}_{\bf{k}}A_{01k}+a^{(3)}_{\bf{k}}A_{02k})
    e^{i\bf{k}\cdot \bf{x}}+h.c.\Big] ,  \label{N25}
\\
\pi^0_A & = \int\frac{d^3k}{(2\pi)^{\frac32}}
\Big(\big(a_{\bf{k}}^{(0)} \pi_{A\, 1k}^{0}
  +a_{\bf{k}}^{(3)}\pi_{A\, 2k}^{0}\big)
          e^{i\bf{k}\cdot\bf{x}}+h.c.\Big) .
   \label{N31}
\el
Both the operators $\pi_A$ and $\pi_A^0$
are expanded by {\small$(a^{(0)}_{\bf{k}}, a^{(0)\dag}_{\bf{k}}, a^{(3)}_{\bf{k}}, a^{(3)\dag}_{\bf{k}})$ }for the Maxwell field, otherwise, the
inconsistencies would arise, such as an imaginary spectral energy density.
 In contrast, for the Stueckelberg field \cite{FrobHiguchi2014,ChuKoyama2017},
$\pi_A^0$ is expanded in terms of
 {\small$(a^{(0)}_{\bf{k}}, a^{(0)\dag}_{\bf{k}})$} and, respectively,
$\pi_A$ in terms of {\small$(a^{(3)}_{\bf{k}}, a^{(3)\dag}_{\bf{k}})$}.

The equal-time covariant canonical commutation relations are imposed
\bl
& [A_\mu(\tau, {\bf x}),\pi^{\nu}_A(\tau, {\bf x'})]
           =i g^{\nu}_{~\mu} \delta({\bf x-x'}),
  \label{N36}
\el
where the creation and annihilation operators satisfy
the covariant commutators
\bl
[a^{(\mu)}_{\bf{k}},a^{(\nu)\dag}_{\bf{k}'}]
& =\eta^{\mu\nu} \delta^{(3)} (\bf k-k') .
 \label{commaamunu}
\el
The transverse modes $f^{(\sigma)}_k(\tau)$
are  given by Eq.\eqref{solutionoftranseverse},
the transverse  polarizations $\epsilon^\sigma_i(k)$
satisfy the usual orthonormal relations \cite{ZhangYePRD2022},
and the  longitudinal  and temporal  $k$-modes are
\bl
A_{1k}&=c_1\frac{1}{a(\tau)}\frac{i}{k}\frac{1 }{\sqrt{2k}}e^{-ik\tau}
+\alpha_1\frac1k\frac{(3+6ik\tau)(\zeta +5)
-6k^2\tau^2(\zeta +1)
-4ik^3\tau^3(\zeta -1)}{24  k\tau}
\frac{1}{\sqrt{2k}}e^{-i k\tau},\label{zz7radd}
\\
A_{2k}&=c_2\frac{1}{a(\tau)}\frac{i}{k}\frac{1 }{\sqrt{2k}}e^{-ik\tau}
+\alpha_2\frac1k\frac{(3+6ik\tau)(\zeta +5)
-6k^2\tau^2(\zeta +1)-4ik^3\tau^3(\zeta -1)}{24 k\tau } \frac{1}{\sqrt{2k}}e^{-i k\tau},\label{zz8rad}
\\
A_{01k}&=c_1\frac{1}{a(\tau)}\frac{1 }{\sqrt{2k}}(1-\frac{i}{k\tau})e^{-ik\tau}
+\alpha_1\frac{ 4 k^4 \tau ^4 (1-\zeta )
- (2 i k^3 \tau ^3+3i k \tau +3) (\zeta +5)}{24   k^2 \tau^2 }
 \frac{1}{\sqrt{2k}} e^{-i k \tau } ,\label{zz5rad}
\\
A_{02k}&=c_2\frac{1}{a(\tau)}\frac{1 }{\sqrt{2k}}(1-\frac{i}{k\tau})e^{-ik\tau}
+\alpha_2\frac{ 4 k^4 \tau ^4 (1-\zeta )
- (2 i k^3 \tau ^3+3i k \tau +3) (\zeta +5)}{24   k^2 \tau^2 }
\frac{1}{\sqrt{2k}} e^{-i k \tau },\label{zz6rad}
\el
\bl
\pi_{A\, 1k}^{0} & =  \alpha_1\frac{a(\tau)}{a_r} k^2 \frac{1}{\sqrt{2k}}
  e^{-i k \tau },\label{zz9}
\\
\pi_{A\, 2k}^{0} & = \alpha_2  \frac{a(\tau)}{a_r} k^2  \frac{1}{\sqrt{2k}}
         e^{-i k \tau },
        \label{zz10}
\\
\pi_{A\, 1k} &   = \alpha_1\frac{a(\tau)}{a_r} \, i k (1-\frac{i}{k\tau})
     \frac{1}{\sqrt{2k} } e^{-i k \tau },\label{zz11}
\\
\pi_{A\, 2k} & = \alpha_2\frac{a(\tau) }{a_r} \, i k (1-\frac{i}{k\tau})
       \frac{1}{\sqrt{2k} } e^{-i k \tau }, \label{zz12}
\el
where  $(c_1, \alpha_1), (c_2,  \alpha_2)$
are two sets of  dimensionless complex coefficients,
subject to the following constraints due to
the commutation relations \eqref{N36} and \eqref{commaamunu}
\bl
&|c_1|^2-|c_2|^2=0,\label{c2m2constraintrad}
\\
&|\alpha_1|^2-|\alpha_2|^2=0,\label{c1m1constraintrad}
\\
&c_2\alpha_2^*-c_1\alpha_1^*
   = i   \frac{a_r}{k}  \,  . \label{c2c1c2m1cconstraintrad}
\el
Obviously,
$c_1\ne 0, c_2\ne 0, \alpha_1\ne 0, \alpha_2\ne 0$,
ie,  both the homogeneous and inhomogeneous parts of $A$ and $A_0$
must be present for a consistent covariant canonical quantization.
A simple choice satisfying
 \eqref{c2m2constraintrad}--\eqref{c2c1c2m1cconstraintrad}  is
\bl
& c_1=c_2=1,
  ~~~~ \alpha_1  =  - \alpha_2 =  i \frac{a_{r}}{2 k} \,  .
     \label{allowc12rad}
\el
Under the residual gauge transformations  \eqref{Atrsfrad} and \eqref{A0trsfrad},
only the homogeneous parts  of $A$ and $A_0$ shift as
\bl
& c_1 \rightarrow c'_1=c_1 +C_1,
  \label{c1c}
 \\
&  c_2 \rightarrow c'_2=c_2 +C_2,
\label{c2c}
\el
where $C_1$ and $C_2$ are two complex constants.
In analogy  to \eqref{c2m2constraintrad}--\eqref{c2c1c2m1cconstraintrad},
the new coefficients also obey the constraints
\bl
&|c_1'|^2-|c_2'|^2=0,\label{c2m2constraintradNew}
\\
&|\alpha_1|^2-|\alpha_2|^2=0,\label{c1m1constraintradNew}
\\
&c_2'\alpha_2^*-c_1'\alpha_1^*=i \frac{a_{r}}{k} ,
   \label{c2c1c2m1cconstraintradNew}
\el
which lead to  the following restrictions
\bl
&  |C_1|^2-|C_2|^2+ 2Re(c_1C_1^{*}-c_2 C_2^{*})   =0,
 \label{cetrans}
\\
&  \alpha_2^* C_2-\alpha_1^* C_1  =0. \label{forgaugetrans}
\el
For the coefficient choice \eqref{allowc12rad},
the solutions of  \eqref{cetrans}  and \eqref{forgaugetrans} are
\bl
C_1   = - C_2  = i \, r ,
  \label{reggtf}
\el
with  $r$ being an arbitrary real number.
As a result, the nonvanishing transformed homogeneous parts
are ensured
$c'_1     = 1 + i r \ne 0$,  $c'_2    = 1- i r \ne 0$.
The transformation prescribed by \eqref{reggtf}
is a subset of the transformation
\eqref{Atrsfrad} and \eqref{A0trsfrad} at the classical level,
and  is referred to as
the quantum residual gauge transformation \cite{ZhangYePRD2022}.

\section{The stress tensor  in RD stage}

Analogously to  de Sitter space \cite{ZhangYePRD2022},
the stress tensor of the Maxwell field with the GF  term
in the RD stage   consists of three parts:
\bl \label{rhofl}
\rho   & = - T^0\,_0 = \rho^{TR} + \rho^{LT} + \rho^{GF} ,
          \\
 p  &  = \frac13 T^j\, _j  = p^{TR} + p^{LT} + p^{GF} .
 \label{pfl}
\el
The transverse and LT  stress tensor are
\bl
\rho^{TR} & =  3 p^{TR}
= \frac{1}{2} a^{-4} \Big( B_j ' B_j' + B_{i,j} B_{i,j}  \Big)  ,
\\
\rho^{LT} &  = 3p^{LT} =\frac{1}{2} a^{-4}
      \partial_i\pi_A \partial^i \pi_A  \, ,
  \label{longittempml}
\el
both being  gauge invariant and independent of $\zeta$.
The GF  stress tensor due to the GF term is
\bl
\rho^{GF}  & =   \frac{1}{a^4} \Big[ - \frac{1}{2} \zeta  ( \pi^0_A  )^2
      - A_0 \Big(  \partial_0 \pi^{0}_A   - D \pi^0_A  \Big)
      - A_{,j} \pi^0_{A \, ,j}   \Big]  ,     \label{rhoGFml}
  \\
p^{GF}  & =  \frac{1}{a^4 }
  \Big[\frac{1}{2} \zeta (  \pi^0_A  )^2
  - A_0 \Big(  \partial_0 \pi^{0}_A   - D \pi^0_A  \Big)
  + \frac13 A_{,j} \pi^0_{A \, ,j}  \Big] ,  \label{GFpreml}
\el
which apparently depends on  $\zeta$
and varies  under the residual gauge transformation.

The expectation value of the stress tensor
are the source of Einstein equation \cite{DeWitt1975,DeWittBrehme1960}.
The transverse  stress tensor
in a transverse state $|\phi\rangle$  is
\bl \label{rhomasslessdeSitter}
\langle \phi|  \rho^{TR} |\phi \rangle
 = & 3 \langle \phi|p^{TR} |\phi \rangle
=   \int^{\infty}_0   \rho^{TR}_k   \frac{d k}{k}
  + \int\frac{d k}{k} \rho^{TR}_k
   \sum_{\sigma=1,2}
  \langle \phi|a_{\bf k}^{(\sigma)\dag} a_{\bf k}^{(\sigma)} |\phi\rangle,
\el
where the first term is the vacuum part
with  the spectral energy density
\bl \label{rhopTRvac}
\rho^{TR}_k  =  3 p^{TR}_k  &
 = \frac{k^3}{2\pi^2 a^4}
  \Big[ |f_k^{(1)'}(\tau)|^2 + k^2 |f_k^{(1)}(\tau)|^2  \Big]
  = \frac{k^4}{2 \pi^2 a^4}   ,
\el
and the second term of \eqref{rhomasslessdeSitter}
is the  photon part.
For photons in thermal equilibrium during the RD stage,
the photon number distribution is
$ \langle \phi|a_{\bf k}^{(\sigma)\dag} a_{\bf k}^{(\sigma)} |\phi\rangle
= 1/(e^{k/T}-1)$
and the photon energy density is
$\int\frac{d k}{k} \rho^{TR}_k \sum_{\sigma}
 \langle \phi|a_{\bf k}^{(\sigma)\dag} a_{\bf k}^{(\sigma)} |\phi\rangle
 = \frac{\pi^2}{15}(\frac{T}{a(\tau)})^4$.
In this paper we are more interested in the vacuum part.

The GB  physical states $|\psi\rangle$
\cite{Gupta1950,Gupta1977,ItzyksonZuber,ZhangYePRD2022}
for the longitudinal and temporal fields are defined such that
the positive frequency part of
the temporal  canonical momentum operator  annihilates the state,
$\pi^{0(+)}_A|\psi\rangle =0$,
ie,
\bl
( \alpha_1a^{(0)}_{\bf k}+\alpha_2a^{(3)}_{\bf k} )|\psi\rangle=0 ,
\label{gbrelation}
\el
which is
$(a^{(0)}_{\bf k} - a^{(3)}_{\bf k})|\psi\rangle=0$
for the choice  $\alpha_1=-\alpha_2$ of \eqref{allowc12rad},
and also implies
\bl  \label{LTGBcondition}
\langle\psi|a^{(0)\dag}_{\bf k} a^{(0)}_{\bf k} |\psi\rangle
   =\langle\psi|a^{(3)\dag}_{\bf k} a^{(3)}_{\bf k} |\psi\rangle .
\el
The GB condition \eqref{gbrelation} also implies that $\pi^{(+)}_A|\psi\rangle =0$,
as one can see through the expression \eqref{N33}.
(In  the b-field formalism  \cite{Ojima1990},
the physical states introduced to remove the influences of
the auxiliary b-field will lead to
the same condition  \eqref{gbrelation} of the GB states.)
During the whole RD stage,
the positive frequency modes \eqref{Arsolution4th}, \eqref{A0rsolution4th},
\eqref{pir} and  \eqref{pir0} remain $\propto e^{-ik\tau}$
and will not change to the negative frequency modes $\propto e^{-ik\tau}$,
so that the GB condition \eqref{gbrelation} holds consistently
\cite{Higuchi1990}.
(Analogously, for the MD stage the GB condition can be imposed.)
The expectation of the LT stress tensor in the  GB physical state is
\bl
\langle\psi|\rho^{LT}|\psi\rangle
& = 3 \langle\psi|p^{LT}|\psi\rangle
=      \int\rho^{LT}_{k} \frac{d k}{k } ,
   \label{rhoLT}
\el
where
\bl
\rho^{LT}_{k}
&=  \frac{k^5}{4 \pi^2  a^4}
 \bigg(2\langle\psi| a^{(0)\dag}_{\bf k} a^{(0)}_{\bf k}
 |\psi\rangle |\pi_{A\, 1k}|^2
+2\langle\psi| a^{(3)\dag}_{\bf k} a^{(3)}_{\bf k}  |\psi\rangle |\pi_{A\, 2k}|^2
  - |\pi_{A\, 1k}|^2   + |\pi_{A\, 2k}|^2 \bigg)
 \nn \\
& +  \frac{k^5}{4 \pi^2  a^4}
 \bigg(   2 \langle\psi| a^{(3)\dag}_{\bf k} a^{(0)}_{\bf k}
  |\psi \rangle \pi_{A\, 2k}^*\pi_{A\, 1k}
 + 2 \langle\psi| a^{(0)\dag}_{\bf k} a^{(3)}_{\bf k}
  |\psi\rangle  \pi_{A\, 1k}^*\pi_{A\, 2k}    \bigg)
\nn \\
& +   \frac{k^5}{4 \pi^2  a^4}
\bigg(\langle\psi|a^{(0)}_{\bf k} a^{(0)}_{-\bf k} |\psi\rangle \pi_{A\, 1k}^2
+\langle\psi|a^{(3)}_{\bf k} a^{(0)}_{-\bf k} |\psi\rangle
 \pi_{A\, 2k}\pi_{A\, 1k}
\nn \\
& ~~~~~~~~~~
 +\langle\psi|a^{(0)}_{\bf k} a^{(3)}_{-\bf k} |\psi\rangle
 \pi_{A\, 1k}\pi_{A\, 2k}
+\langle\psi|a^{(3)}_{\bf k} a^{(3)}_{-\bf k}|\psi\rangle
\pi_{A\, 2k}^2
  +h.c.\bigg).\label{LTrho}
\el
Applying  the GB condition \eqref{gbrelation} and \eqref{LTGBcondition},
 the longitudinal and  temporal parts cancel each other,
\bl
\rho_k^{LT} = 3p_k^{LT} =0,\label{masslessLT}
\el
including both the particle and vacuum parts.
The vacuum part cancels out even without
the GB condition \eqref{LTGBcondition}.
So the LT stress tensor in the GB state
is zero even before regularization.
This is also true  in de Sitter space \cite{ZhangYePRD2022}
and in the MD stage (see Appendix \ref{maxwllmds}) as well.

The GF  stress tensor  in the GB physical state is
\bl
\langle \psi |\rho^{GF }|\psi\rangle
 & =   \int\rho^{GF}_{k} \frac{d k}{k }  ,\label{gfPIPI}
  \\
\langle \psi |p^{GF }|\psi\rangle
& =  \int p^{GF}_{k} \frac{d k}{k } ,
  \label{PIPIPI}
\el
with the GF spectral energy density  and pressure given by
\bl
\rho_{k}^{GF}=\frac{k^4}{2\pi^2a^4}
\Big( \langle\psi| a_{\bf k}^{(3)\dag} a_{\bf k}^{(3)} |\psi\rangle
 -\langle\psi|a_{\bf k}^{(0)\dag}a_{\bf k}^{(0)}|\psi\rangle
 +1 \Big)
   \Big(1+\frac{1}{2 k^2 \tau ^2} \Big),
\label{particlerhogfrad}
\el
\bl
p_{k}^{GF}=\frac{k^4}{2\pi^2a^4}
\Big(\langle\psi|a_{\bf k}^{(3)\dag}a_{\bf k}^{(3)}|\psi\rangle
 -\langle\psi|a_{\bf k}^{(0)\dag}a_{\bf k}^{(0)}|\psi\rangle
 +1 \Big)
\frac13 \Big(1+\frac{3}{2 k^2 \tau ^2} \Big) ,
\label{particlepgfrad}
\el
which are independent of the gauge fixing constant $\zeta$,
and contributed only by the homogeneous part of $k$ modes $A_{0k}$ and $A_k$.
It is seen that the total stress tenor of Maxwell field
in the GB state is independent of $\zeta$.
This property is interesting,
 and also consistent with a more generic  result that
the physical quantities should be independent of the gauge fixing constant
in the context for Yang-Mills theories \cite{Piguet1985}.
By the GB condition \eqref{LTGBcondition},
the particle part of the GF stress tensor  cancels out,
$\langle\psi|a_{\bf k}^{(3)\dag}a_{\bf k}^{(3)}|\psi\rangle
-\langle\psi|a_{\bf k}^{(0)\dag}a_{\bf k}^{(0)}|\psi\rangle=0$,
and only the vacuum part remains
\bl
\rho_{k}^{GF} =\frac{k^4}{2\pi^2a^4}(1+\frac{1}{2k^2\tau ^2}) ,
 \label{rhogfrad}
\\
p_{k}^{GF}=\frac{k^4}{2\pi^2a^4}\frac{1}{3}( 1+\frac{3}{2k^2\tau ^2}),
\label{pgfrad}
\el
which is twice of that of the minimally coupling massless scalar field
in the RD stage\cite{ZhangWangYe2020}.
The expression \eqref{rhogfrad} of $\rho_{k}^{GF}$
has the same form as  that in de Sitter space,
but the expression \eqref{pgfrad} of $p_k^{GF}$
differs from that in de Sitter space \cite{ZhangYePRD2022}.
If a ghost field was introduced \cite{AdlerLiebermanNg1977,BrownCassidy1977},
the particle part of the ghost stress tensor
has nothing to cancel and is unaccountable,
since the particle part of the GF stress tensor
\eqref{particlerhogfrad}  and \eqref{particlepgfrad} is zero.

Under the quantum residual gauge transformation
 \eqref{c1c}$\sim$\eqref{c2c} with \eqref{reggtf},
the GF stress tensor \eqref{rhogfrad} and \eqref{pgfrad} varies to
\bl
&\delta\rho_{k}^{GF}
=\frac{k^3}{2\pi^2a^4}\frac{i(\alpha_1^* C_1 -\alpha_2^* C_2 )k^2 }{a_{r}}
  \Big(  \langle\psi|a_{\bf k}^{(3)\dag}a_{\bf k}^{(3)}|\psi\rangle
-\langle\psi|a_{\bf k}^{(0)\dag}a_{\bf k}^{(0)}|\psi\rangle + 1 \Big)
   \Big(1+\frac{1}{2 k^2 \tau^2 }  \Big),\label{vaiationrho}
\el
\bl
\delta p^{GF}_{k}&=\frac{k^3}{2\pi^2a^4}
\frac{i(\alpha_1^* C_1 -\alpha_2^* C_2 ) k^2}{a_{r}}
  \Big( \langle\psi|a_{\bf k}^{(3)\dag}a_{\bf k}^{(3)}|\psi\rangle
  -\langle\psi|a_{\bf k }^{(0)\dag}a_{ \bf k }^{(0)}|\psi\rangle +1  \Big)
\frac13  \Big(1+\frac{3}{2 k^2 \tau^2 }  \Big),\label{vaiationp}
\el
which is zero,
$\delta\rho_{k}^{GF}=0$, $\delta p_{k}^{GF}=0$ by
the restriction \eqref{forgaugetrans}.
Thus, the GF tress tensor is invariant.

\section{The regularization of stress tensor in   RD  stage}

The LT stress tensor is already zero  in the GB state.
The transverse and  GF vacuum stress tensors are UV divergent,
and need to be regularized as the following.

The transverse \eqref{rhopTRvac}
has  one divergent term $\propto k^4$,
and the 0th-order adiabatic regularization is sufficient.
By analogous calculations to those in Ref.\cite{ZhangYePRD2022},
the 0th-order adiabatic transverse modes are
the same as the exact transverse modes
\bl
 f_{k\, 0th}  =\frac{1}{\sqrt{2 k}}e^{- ik \tau}
= f_k^{(\sigma)} , ~~~ (\sigma=1,2) \label{ad0f}
\el
[Actually the transverse modes of all adiabatic orders are equal to
the exact modes, $f_{k\, 0th}  = f_{k\, 2nd}  = ... = f_k^{(\sigma)}$,
like a conformally-coupling massless scalar field
\cite{ZhangYeWang2020,ZhangWangYe2020}.]
As a result,
the 0th-order subtraction term for the vacuum transverse
spectral stress tensor  is
\bl \label{adrrhtr}
\rho^{TR }_{k \, \, 0th}  & = 3  p^{TR}_{k  \, \, 0th}  =
  \frac{k^3}{2\pi^2 a^4}
  \Big[ |f_{k\, 0th}'(\tau)|^2 + k^2 |f_{k\, 0th}(\tau)|^2  \Big]
    = \rho^{TR}_{k}  = 3p^{TR}_{k} ,
\el
equal to the exact  \eqref{rhopTRvac}.
So, the 0th-order regularized transverse vacuum
spectral stress tensor is
\bl \label{adrrhtrho}
\rho^{TR}_{k\, reg} \equiv \rho^{TR}_{k} -\rho^{TR }_{k \, \, 0th}   =0,
\\
p^{TR}_{k\, reg} \equiv p^{TR}_{k} - p^{TR }_{k \, \, 0th}  =0 ,
  \label{adrrhtp}
\el
which is true for all the fRW spacetimes.

The GF  vacuum stress tensor \eqref{rhogfrad} and \eqref{pgfrad}
in the RD stage  contains  $k^4,k^2$ divergences,
and the 2nd-order adiabatic regularization is sufficient.
[In contrast, for the MD stage
 the GF  vacuum stress tensor contains $k^4,k^2, k^0$ divergences,
and the 4th order regularization is needed, as we shall see in Appendix.]
The equation \eqref{equationofpiA0res} of rescaled $\bar \pi^{0}_A $
is  the same as the equation of a comformal-coupling massless scalar field
 \cite{ZhangYeWang2020,ZhangWangYe2020},
so the modes of all adiabatic order are equal
\bl
\bar \pi_{A\,   0th}^{0} & = \bar \pi_{A\,   2nd}^{0} = ... = \frac{1}{\sqrt{2 k} }
  e^{-i k \tau} ,
\el
the same as the exact mode  \eqref{pi0rpi0rbar}.
Multiplying by $a(\tau)$ gives  $\pi_{A\,  2nd}^{0}=\pi^{0}_A$,
ie,  the 2nd order adiabatic mode is equal to
the exact modes \eqref{zz9}\eqref{zz10}.
Since only the homogeneous part of  $A$ and $A_0$
contribute to the GF stress tensor,
we just need the 2nd-order adiabatic modes of  the homogeneous parts.
By analogous calculations to those in Ref.\cite{ZhangYePRD2022},
the 2nd-order adiabatic modes  are found to be
the same as  the homogeneous parts of \eqref{zz7radd} -- \eqref{zz6rad},
and  the second-order  subtraction terms
for the  vacuum GF spectral stress tensor  are
\bl  \label{GFrhosub}
\rho^{GF}_{k\, 2nd}
  & = \frac{k^4}{2\pi^2  a^4} \Big(1 + \frac{1}{2 k^2 \tau^2} \Big)
    = \rho^{GF}_{k},
\\
p^{GF}_{k\, 2nd}
  & =  \frac{1}{3} \frac{k^4 }{ 2\pi^2   a^4}
  \Big( 1 + \frac{3}{2 k^2 \tau^2} \Big) = p^{GF}_{k},
  \label{GFprsub}
\el
just equal to the exact expressions  \eqref{rhogfrad} and \eqref{pgfrad}.
So the second-order regularized vacuum GF stress tensor is zero,
\bl
\rho^{GF}_{k\, reg} \equiv \rho^{GF}_{k} -\rho^{GF}_{k\, 2nd} =0 ,
   \label{GFrhoreg}
\\
p^{GF}_{k\, reg} \equiv p^{GF}_{k} - p^{GF}_{k\, 2nd} =0 .
           \label{GFpreg}
\el
(Obviously, the subtraction stress tensor
corresponding to \eqref{adrrhtr}, \eqref{GFrhosub} and \eqref{GFprsub},
 is covariant, and also covariantly conserved.)
Hence, in the RD stage, just as in de Sitter space \cite{ZhangYePRD2022},
the   regularized GF stress tensor is vanishing,
and the total regularized vacuum stress tensor of
the Maxwell field is zero \cite{UtiyamaDeWitt1962,AdlerLiebermanNg1977},
and there is no trace anomaly.
This also confirms
the result \cite{ZhangYePRD2022, ZhangYeWang2020, ZhangWangYe2020, YeZhangWang2022}
that the trace anomaly will not arise when one deals directly with
a massless field from the beginning.
In literature\cite{BrownCassidy1977,DowkerCritchley1977,ChuKoyama2017},
the trace anomaly came up as an artifact
from the following consecutive procedures:
starting with a massive vector field,
applying the 4th order regularization on the stress tensor
to remove the UV divergences,
and finally taking the massless limit of the trace of
the regularized stress tensor of the massive field.
However, as we pointed out
\cite{ZhangYePRD2022,ZhangYeWang2020, ZhangWangYe2020, YeZhangWang2022},
the 4th order regularization is improper
for the stress tensor of
the conformally- and minimally-coupling scalar fields,
and here it will be improper for the Maxwell field in the RD stage either.
This is because, as we find for the above fields,
the 4th order regularization
violates the minimal subtraction rule \cite{ParkerFulling1974},
leads to an unphysical,  negative spectral energy density,
and as well as causes singularities of the regularized stress tensor
in the massless limit.
These three vital shortcomings have not been noticed nor examined
by the literatures that claimed the trace anomaly.
Ref. \cite{Glavan2212.13982} used the operator ordering
and also obtained the zero GF stress tensor as
\eqref{GFrhoreg} and \eqref{GFpreg}.
However, the operator ordering is actually
the normal ordering that simply drops the zero-point energy,
and thus is an improper scheme for quantum fields in curved spacetimes,
as has long been emphasized in Refs.\cite{FeynmanHibbs1965,DeWitt1975}.
Our result tells that the regularized vacuum stress tensor of the Maxwell field
is  vanishing and can not contribute to the cosmological constant
that was suggested by Refs.\cite{JimenezMaroto2009,JimenezMaroto2010}.
Instead,  a massive scalar field has a nonzero regularized vacuum stress tensor
which can contributes to the cosmological constant
\cite{ZhangYeWang2020,YeZhangWang2022}.
Since both the particle and vacuum parts of the GF stress tensor are zero,
one needs no introduction \cite{AdlerLiebermanNg1977,BrownCassidy1977}
 of a ghost field at all.
Ref.\cite{Vollick2012} used  the normal ordering
to give a zero vacuum stress tensor
of the Maxwell field in the Lorenz gauge ($\nabla^{\mu}A_{\mu}=0$).
But  the normal ordering is not an appropriate procedure
to use in curved spacetimes,
and, moreover,
the Lorenz gauge  amounts to dropping the GF term and
the covariance canonical quantization will be lost.

\section{Conclusion and Discussion}

We study  the Maxwell field with a general GF term
in the respective RD and  MD stages,
and extend our pervious work
in de Sitter space and in Minkowski spacetime \cite{ZhangYePRD2022}.
The main procedures, such as the derivation of the solutions,
the implementation of covariant canonical quantization,
the calculation of the stress tensor,
and the adiabatic regularization of the vacuum stress tensor,
are similar in the RD and MD stages,
and are also analogous to those in de Sitter space \cite{ZhangYePRD2022}.
Nevertheless,   the order of regularization for
the GF stress tensor for  the MD stage differs
from that for the RD and de Sitter stages.

The transverse solution \eqref{solutionoftranseverse} is simple,
and  holds for all fRW spacetimes \cite{ZhangYePRD2022}.
The nontrivial part is
the longitudinal and temporal components $A$ and $A_0$,
as   their equations get  mixed up due to the  GF term.
In two different ways, we have been able to derive their analytical solutions
\eqref{Arsolution4th} and \eqref{A0rsolution4th} for the RD stage,
as well as  \eqref{Arsolutionma} and \eqref{A0rsolutionma} for the MD stage.
Both $A$ and $A_0$ are composed of
the homogeneous and inhomogeneous solutions,
and the canonical momenta  $\pi_A$ and $\pi_A^0$
are contributed by the inhomogeneous pieces only.
For a consistent covariant canonical quantization,
both the homogeneous and inhomogeneous $k$-modes
must be present in the operators $A$ and $A_0$,
and, moreover, the residual gauge transformations
are restricted to the quantum residual gauge transformations
prescribed by \eqref{reggtf}  as a subset.

The structure and regularization of the stress tensor
 of the Maxwell field with the GF  term are particularly revealing.
The transverse stress tensor has both the particle and vacuum parts,
and the vacuum part
contains only a UV divergent  $k^4$ term,
which becomes zero after the 0th-order adiabatic regularization.
The LT stress tensor is zero in the GB states,
including the particle and vacuum,
due to a cancelation of the longitudinal and temporal parts.
The particle part of the GF stress tensor is also zero in the GB states.
The above features  hold in the RD and MD stages,
and in de Sitter space as well.
But, the vacuum GF stress tensor exhibits different behaviors
in the RD and MD stages:
In the RD stage it contains two  divergent terms
$\propto k^4, k^2$ as in \eqref{rhogfrad} and \eqref{pgfrad},
and becomes zero by the 2nd-order adiabatic regularization,
 nevertheless, in the MD stage,
it contains three  divergent terms $\propto k^4, k^2, k^0$
as in \eqref{rhogfmat} and \eqref{pgfmat}
and becomes zero by the 4th-order adiabatic regularization
\cite{ParkerFulling1974,WangZhangChen2016,ZhangWangJCAP2018}.
This outcome tells that
the adiabatic order of an adequate regularization  generally
depends upon the background spacetimes,
the type of fields (or the components of vector field),
as well as the  couplings
\cite{ZhangYeWang2020,ZhangWangYe2020,YeZhangWang2022,ZhangYePRD2022}.

Hence,  in both the RD and  MD stages,
the total regularized vacuum stress tensor is zero
for the Maxwell field with a general GF term,
only the photon part of transverse stress tensor remains.
In particular, our  result demonstrates
that the vanishing GF stress tensor
can not be a candidate for the dark energy,
and that there is no trace anomaly,
disproving the claim in literature.
We also point out that
the trace anomaly came up as an artifact from  the improper use
 of the 4th-order regularization
 on some strangely-assumed massive fields,
and that the trace anomaly  will never occur
when one works directly with massless fields.
As an important feature,
the stress tensor of the Maxwell field in the GB states is
independent of the GF constant,
as well as invariant under the quantum gauge transformation.
It should be emphasized that,
since the GF stress tensor is zero,
including both the particle and vacuum parts,
one needs no introduction of a ghost field to cancel
the GF stress tensor,
otherwise the ghost stress tensor itself
would become unaccountable and cause unphysical consequences.
The ghost field can be useful in other topics and for other purposes,
which are beyond our current paper.

\

{\textbf{Acknowledgements}}

Y. Zhang is supported by
NSFC Grant No. 11675165, 11961131007, 12261131497,
 and in part by National Key RD Program of China (2021YFC2203100).

\appendix
\numberwithin{equation}{section}

\section{Maxwell field with GF term in  MD stage}\label{maxwllmds}

In this Appendix,  we list  the main results in the MD stage,
all the procedures are analogous to the RD stage.
The scale factor of the MD stage is
\be
 a=a_m \tau^2,
\ee
where $a_m$ is a constant.
In place of \eqref{4th0rdersolutionA} and \eqref{4th0rdersolutionA0},
the fourth-order differential equations of $A$ and $A_0$ in  the MD stage
are
\bl
&[-(\zeta -1)^2 k^2 \tau ^2+4 (5 \zeta -9)]\tau ^4  A^{(4)}(\tau )
                        + 8 (5 \zeta -9)  \tau^3  A^{(3)}(\tau )
\nn
\\
&~~~~-2 [(\zeta -1)^2 k^4 \tau ^4-2  k^2 \tau ^2 (\zeta (5 \zeta  +2)-15)
                            +40 (5\zeta-9)] \tau^2 A''(\tau )
  -8 (\zeta -2) k^2 \tau ^2 (5 \zeta   -9)  \tau   A'(\tau )
\nn
\\
&~~~~-k^2 \tau ^2 [16 (5\zeta-9)+(\zeta -1)^2 k^4 \tau ^4
          +4 k^2 \tau ^2 (\zeta  (\zeta -9)+12)]  A(\tau)
    =0,\label{4th0rdersolutionAmat}
\el
and
\bl
&[(\zeta -1)^2 k^2 \tau ^2+4 \zeta    (\zeta +3)]\tau ^4 A_0^{(4)}(\tau )
   +8 \zeta  (\zeta +3)\tau ^3 A_0^{(3)}(\tau )
\nn
\\
&~~~~~ +2  [-40 \zeta (\zeta +3)+(\zeta -1)^2 k^4 \tau ^4
       + 2 k^2\tau ^2 (\zeta  (-3 \zeta  +18)-7)]\tau ^2 A_0''(\tau )
\nn
\\
&~~~~~ +8    [20  \lambda   (\zeta +3)+k^2 \tau ^2
(\zeta  (12\zeta -15)+7)]\tau A_0'(\tau )\nn
\\
&~~~~~ - [-(\zeta -1)^2k^6\tau ^6+4 k^4 \tau ^4  (-2 \zeta ^2-3 \zeta +1)
     +160 \zeta (\zeta +3)+8 k^2\tau ^2 (\zeta  (13\zeta -12)+7)] A_0(\tau )
   =0,\label{4th0rdersolutionA0mat}
\el
holding  for a general $\zeta$,
and the positive frequency solutions are
\bl
A&= d \frac{1}{a(\tau)}\frac{i}{k}\frac{1 }{\sqrt{2k}}
   \Big(1-\frac{i}{k\tau} \Big) e^{-ik\tau}
\nn
\\
&+\beta\frac1k\frac{  12 i k^6 \tau ^6 (1-\zeta)
-6 k^5 \tau ^5 (7 \zeta+3)+10 i k^4 \tau ^4 (7 \zeta +27)
+\left(70 k^3 \tau ^3+105 (k \tau - i)\right)
(\zeta+9)}{  120 k^3 \tau ^3}  \frac{1}{\sqrt{2 k}}e^{-i k \tau },
\label{Arsolutionma}
\\
A_{0}&= d \frac{1}{a(\tau)}\frac{1}{\sqrt{2 k}}
 \Big(1-\frac{3 i}{k \tau }-\frac{3}{k^2 \tau ^2} \Big) e^{-i k \tau }
\nn
\\
&+\beta\frac{  12 k^7 \tau ^7 (1-\zeta  )
-6 i k^6 \tau ^6 (11-\zeta  )-(14 k^5 \tau ^5
+105 i k^2 \tau ^2+315 (-k \tau +i) )(\zeta  +9) }{120 k^4 \tau ^4}
\frac{1}{\sqrt{2k}}e^{-i k \tau}
\label{A0rsolutionma}
\el
with $d$ and $\beta$ being dimensionless constants.
\eqref{Arsolutionma} and \eqref{A0rsolutionma} satisfy
the basic second-order equations \eqref{equationAA02} and \eqref{equationAA01}.
The corresponding  canonical momentums are
\bl
\pi_A &=\beta  k^3
 \frac{i}{k}\frac{a(\tau)}{a_{m}}\frac{1}{\sqrt{2 k}}
   \Big(1-\frac{3 i}{k \tau }
   -\frac{3}{k^2 \tau ^2}\Big) e^{-i k \tau },\label{pirmat}
\\
\pi_A ^{0}&=\beta  k^3  \frac{a(\tau)}{a_{m}}\frac{1 }{\sqrt{2k}}
   \Big(1-\frac{i}{k\tau}\Big) e^{-ik\tau} ,
\label{pir0mat}
\el
contributed only by the $\beta$ parts of $A$ and $A_{0}$.
The solutions \eqref{Arsolutionma} -- \eqref{pir0mat}
in the  MD stage can
be obtained in another way.
By resealing
$\pi_A =a\bar{\pi}_A$ and  $\pi_A^0=a\bar{\pi}^0_A$ with
the scale factor $a=a_m \tau^2$,
Eqs.\eqref{equationofpiA} and \eqref{equationofpiA0} reduce to
\bl
\partial_0^2\bar{\pi}_A +  (k^2-\frac{6}{\tau ^2})\bar{\pi}_A
=0\label{equationofpiAresraidma},
\\
\partial_0^2\bar{\pi}^{0}_A +  (k^2-\frac{2}{\tau ^2})\bar{\pi}^{0}_A
=0\label{equationofpiA0resraidma},
\el
the normalized solutions are
\bl
\bar{\pi}_A &=\beta k^3  \frac{i}{k}\frac{1}{a_{m}}
\frac{1}{\sqrt{2 k}}\Big(1-\frac{3 i}{k \tau }
   -\frac{3}{k^2 \tau ^2}\Big) e^{-i k \tau },
\\
\bar{\pi}^{0}_A &=\beta k^3  \frac{1}{a_{m}}\frac{1 }{\sqrt{2k}}
   \Big(1-\frac{i}{k\tau}\Big) e^{-ik\tau} .\label{pi0mma}
\el
Multiplying the above by $a(\tau)$ gives the solutions
\eqref{pirmat} and \eqref{pir0mat}.
The  inhomogeneous equations \eqref{AEu2} and \eqref{A0epi}
hold also in the MD stage,
and by use of the given  $\pi_A$ and $\pi_A^0$
of \eqref{pirmat}  and  \eqref{pir0mat},
the solutions are given by \eqref{Arsolutionma}  and \eqref{A0rsolutionma},
where the $d$ parts are the general homogeneous solutions
and the $\beta $ parts are the inhomogeneous solutions.

In the MD stage,
 in place of \eqref{soltheta},
the residual gauge transformation is prescribed  by
\bl
\theta_k
=  M  \frac{1}{a(\tau)} \frac{i}{k}\frac{1 }{\sqrt{2k}}
\Big(1-\frac{i}{k\tau}\Big)e^{-ik\tau}
\label{solthetamat}
\el
with $M$ being a complex constant,
the transverse fields and  the canonical momenta are also invariant,
$B_i \rightarrow   B_i$,
$\pi_A  \rightarrow  \pi_A $, $ \pi^{0}_A  \rightarrow  \pi^{0}_A$,
and,
the longitudinal and temporal $k$-modes transform as
\bl
A_{k} &  \rightarrow
A_k+ M\frac{1}{a}\frac{i}{k}\frac{1 }{\sqrt{2k}}
\Big(1-\frac{i}{k\tau}\Big)e^{-ik\tau},
\label{Atrsfradmat}
  \\
A_{0\, k} & \rightarrow
A_{0\,k}+M\frac{1}{a}\frac{1}{\sqrt{2 k}}
\Big(1-\frac{3 i}{k \tau }-\frac{3}{k^2 \tau ^2}\Big) e^{-i k \tau },
\label{A0trsfradmat}
\el
which amount to a change of the coefficients of
the homogeneous parts of $A_k$ and $A_{0k}$,
\bl \label{MDgtrf}
d \rightarrow  d' = d+ M.
\el

The covariant canonical quantization in the MD stage is the following.
The quantization of the transverse  $B_i$ and $w^i$
is the same as in the RD stage.
The longitudinal and temporal operators for a
general $\zeta$ are written similar to \eqref{N26}--\eqref{N31}
with the $k$ modes
\bl
A_{1k}&=d_1\frac{1}{a}\frac{i}{k}\frac{1 }{\sqrt{2k}}
(1-\frac{i}{k\tau})e^{-ik\tau}
\nn
\\
&+\beta_1\frac1k\frac{  12 i k^6 \tau ^6
(1-\zeta)-6 k^5 \tau ^5 (7 \zeta+3)
+10 i k^4 \tau ^4 (7 \zeta +27)
+\left(70 k^3 \tau ^3+105 (k \tau - i)\right)
 (\zeta+9)}{  120 k^3 \tau ^3}
 \frac{1}{\sqrt{2 k}}e^{-i k \tau},
 \label{zz7maTt}
\\
A_{2k} & = A_{1k} (\text{with $ d_1\rightarrow d_2$, $\beta_1\rightarrow \beta_2$})
,\label{zz8mat}
\\
A_{01k}&=d_1\frac{1}{a}\frac{1}{\sqrt{2 k}}
(1-\frac{3 i}{k \tau }-\frac{3}{k^2 \tau ^2})
e^{-i k \tau }
\nn
\\
&+\beta_1\frac{  12 k^7 \tau ^7 (1-\zeta  )
-6 i k^6 \tau ^6 (11-\zeta  )-(14 k^5 \tau ^5
+105 i k^2 \tau ^2+315 (-k \tau +i) )(\zeta  +9) }{120 k^4 \tau ^4}
   \frac{1}{\sqrt{2k}}e^{-i k \tau } ,
   \label{zz5mat}
\\
A_{02k}&= A_{01k}  (\text{with $ d_1\rightarrow d_2$, $\beta_1\rightarrow \beta_2$})
, \label{a02mat}
\\
\pi_{A\, 1k} &
   = \beta_1\frac{a}{a_{m}}ik^2\frac{1}{\sqrt{2 k}}(1-\frac{3 i}{k \tau }
   -\frac{3}{k^2 \tau ^2}) e^{-i k \tau },\label{zz11matter}
    \\
\pi_{A\, 2k} & = \frac{\beta_2}{\beta_1} \pi_{1k} ,
\label{zz12matter}
\\
\pi_{A\, 1k}^{0} & =  \beta_1\frac{a}{a_{m}}k^3\frac{1 }{\sqrt{2k}}
   (1-\frac{i}{k\tau})e^{-ik\tau},\label{zz9matter}
\\
\pi_{A\, 2k}^{0} & =  \frac{\beta_2}{\beta_1} \pi_{1k}^{0},\label{zz10matter}
\el
where $(d_1, \beta_1)$, $(d_2, \beta_2)$ are two sets of complex coefficients.
The covariant canonical commutation relations and commutators
are  the same as \eqref{N36}  and \eqref{commaamunu}.
Then we obtain the following constraints upon the coefficients
\bl
&|d_1|^2-|d_2|^2=0,\label{c2m2constraintmat}
\\
&|\beta_1|^2-|\beta_2|^2=0,\label{c1m1constraintmat}
\\
&d_2\beta_2^*-d_1\beta_1^*=i \frac{a_m}{k^2} ,
   \label{c2c1c2m1cconstraintmat}
\el
which can be satisfied by  a simple choice
\be
d_1=d_2=1,
~~~ \beta_1= i\frac{a_{m}}{2k^2},
~~~ \beta_2= - i\frac{a_{m}}{2k^2}.
\ee
Under the residual gauge transformations  \eqref{MDgtrf},
the homogeneous parts  of $A$ and $A_0$ shift as
\bl
& d_1 \rightarrow d'_1=d_1 + M_1,
  \label{c1c}
 \\
&  d_2 \rightarrow d'_2=d_2 + M_2,
\label{c2c}
\el
where $M_1$ and $M_2$ are two complex constants, and are  restricted by
\bl
&  |M_1|^2-|M_2|^2+ 2Re(d_1 M_1^*-d_2 M_2^*)   =0,
 \label{cetransmat}
\\
&  \beta_2^* M_2 -\beta_1^* M_1 =0 .\label{forgaugetransmat}
\el
A simple choice is $M_1=-M_2=i \, r$,  where $r$ is an arbitrary number.
As a result,  the homogenous parts are ensured to be non-vanishing
under the transformation.

The expectation value of the transverse stress tensor during the MD stage is
\bl
\langle \phi|  \rho^{TR} |\phi \rangle
=&   \int^{\infty}_0   \rho^{TR}_k   \frac{d k}{k}
  + \int\frac{d k}{k}   \rho^{TR}_k
   \sum_{\sigma=1,2}
  \langle \phi|a_{\bf k}^{\dag(\sigma)} a_{\bf k}^{(\sigma)} |\phi\rangle,
   \label{rhomasslessdeSitterapp}
\el
\bl
\rho^{TR}_k
 = \frac{k^4}{2 \pi^2 a^4} =   3 p^{TR}_k   . \label{rhopTRvacapp}
\el
The expectation value of the LT stress tensor is zero
\bl
\rho_k^{LT} = 3p_k^{LT} =0,
\el
due to the longitudinal and temporal cancelation in the GB state.
The  expectation value of the GF stress tensor   is
\bl
\rho_{k}^{GF}
&=\frac{k^4}{2\pi^2a^4}\Big[\langle\psi|a_{\vec{k}}^{(3)\dag}
 a_{\vec{k}}^{(3)}|\psi\rangle-
\langle\psi|a_{\vec{k}}^{(0)\dag}a_{\vec{k}}^{(0)}|\psi\rangle
+1 \Big] \Big(1+\frac{2}{k^2\tau^2} +\frac{9}{2k^4\tau^4}\Big) ,
\label{particlerhogfmatapp}
\el
\bl
p_{k}^{GF}
&=\frac{k^4}{2\pi^2a^4} \Big[\langle\psi|a_{\vec{k}}^{(3)\dag}
   a_{\vec{k}}^{(3)}|\psi\rangle
 -\langle\psi|a_{\vec{k}}^{(0)\dag}a_{\vec{k}}^{(0)}|\psi\rangle  +1 \Big]
   \frac{1}{3}\Big(1+\frac{4}{k^2\tau^2}+\frac{27}{2k^4\tau^4}\Big) .
 \label{particlepgfmatapp}
\el
By the GB condition of \eqref{LTGBcondition},
the particle part cancels out, only the vacuum part  remains
\bl
\rho_{k}^{GF} & =\frac{k^4}{2\pi^2a^4}
\Big(1+\frac{2}{k^2\tau^2}   +\frac{9}{2k^4\tau^4}\Big),
\label{rhogfmat}
\\
p_{k}^{GF} & = \frac{k^4}{2\pi^2a^4}\frac{1}{3}
\Big(1+\frac{4}{k^2\tau^2}
+\frac{27}{2k^4\tau^4}\Big),
\label{pgfmat}
\el
which are independent of $\zeta$,
and also invariant under the residual gauge transformations
as can be checked directly.
This is equal to
twice of that of the minimally  coupling massless scalar field
in the MD stage\cite{ZhangWangYe2020}.

The regularization of the vacuum stress tensor in the MD stage
is performed analogously to that in the RD stage.
The transverse vacuum  stress tensor \eqref{rhopTRvacapp}
has  one divergent  $ k^4$ term,
and is removed by the 0th order adiabatic regularization
\bl
\rho^{TR}_{k\, reg} =0= p^{TR}_{k\, reg} .
\el
The GF  vacuum stress tensor \eqref{rhogfmat} and   \eqref{pgfmat}
in the MD stage  contain  $k^4,k^2, k^0$ divergences,
and the 4th-order adiabatic regularization is necessary,
and the 2nd-order regularization would be insufficient.
This is unlike the case of the RD stage.
The equation \eqref{equationofpiA0resraidma} of rescaled $\bar \pi^{0}_A $
is  the same as the equation of
a minimal-coupling massless scalar field in the MD stage
\cite{ZhangYeWang2020,ZhangWangYe2020},
and the WKB solution of \eqref{equationofpiA0resraidma} is
\be
\bar \pi_{A  \,  nth}^{0}
  = (2W(\tau))^{-1/2} \exp \Big[  -i \int^{\tau} W(\tau')d\tau' \Big],\label{Yvntrmat}
\ee
where the effective frequency is
\be
  W(\tau)     = \Big[  \omega^2
-\frac{2}{\tau^2}-\frac12 \left( \frac{ W  '' }{ W}
- \frac32 \big( \frac{W  '}{W} \big)^2 \right)  \Big]^{1/2} ,\label{YequaWktrmat}
\ee
which will be solved iteratively.
The 0th-order is $W_{0th}  = \omega = k $,
and the 2nd-order, 4th-order and all higher order frequencies
 are found to be equal to  the following
\bl
W_{2nd} &    = W_{4th}   = ...=  k -\frac{1}{k \tau^2},
\el
and the 2nd-order, 4th-order and all higher order modes are
\bl
\bar \pi_{A \,  2nd}^{0} & = \bar \pi_{A \,  4th}^{0} = ...
 =  \frac{1}{\sqrt{2 k} }
\big( 1 - \frac{i}{k\tau}   \big)   e^{-i k \tau}
= \bar \pi_{A }^{0} ,
\el
just equal to the exact mode of \eqref{pi0mma}.
Multiplying by $a(\tau)$ yields
 $\pi_{A\,  4th}^{0}=\pi_{A\,  2nd}^{0}=\pi^{0}_A$,
ie,  the 4th-order adiabatic mode
is equal to the exact mode \eqref{pir0mat}.
In analogy to the RD stage,
as only the homogeneous parts of $A$ and $A_0$ contribute to
the GF stress tensor,
we need the 4th-order homogeneous parts of
\eqref{zz7maTt}  and  \eqref{zz8mat}
[also equal to the 2nd order] as the following
\bl
A_{1k } &  =  d_1\frac{1}{a}\frac{i}{k}\frac{1 }{\sqrt{2k}}
(1-\frac{i}{k\tau})e^{-ik\tau} ,
    \label{SSS7mat}
\\
A_{2k }  & =   d_2\frac{1}{a}\frac{i}{k}\frac{1 }{\sqrt{2k}}
(1-\frac{i}{k\tau})e^{-ik\tau},
   \label{SSS9mat}
\el
and the 4th-order  homogeneous parts of
\eqref{zz5mat}  and  \eqref{a02mat}  as the following
\bl
A_{01k }
& = d_1\frac{1}{a}\frac{1}{\sqrt{2 k}}(1-\frac{3 i}{k \tau }
-\frac{3}{k^2 \tau ^2}) e^{-i k \tau } ,\label{SSS8mat}
\\
A_{02k }
& = d_2\frac{1}{a}\frac{1}{\sqrt{2 k}}(1-\frac{3 i}{k \tau }
-\frac{3}{k^2 \tau ^2}) e^{-i k \tau } .\label{SSS10mat}
\el
So the 4th order adiabatic subtraction term for
the GF spectral stress tensor
is equal to the exact \eqref{rhogfmat}  and \eqref{pgfmat}.
Hence  the 4th-order regularized GF vacuum  stress tensor is zero,
\bl
\rho^{GF}_{k\, reg} \equiv \rho^{GF}_{k} -\rho^{GF}_{k\, 4th} =0 ,
   \label{GFrhoregmat}
\\
p^{GF}_{k\, reg} \equiv p^{GF}_{k} - p^{GF}_{k\, 4th} =0 ,
           \label{GFpregmat}
\el
so is the regularized trace
$-\rho^{GF}_{k\, reg} +3 p^{GF}_{k\, reg} =0$.
We notice that, while  the 0th-order regularization
is sufficient for the transverse part,
 the 4th-order regularization is needed
for the GF stress tensor in the MD stage,
unlike the 2nd-order one in the RD stage in Sect 5.
This confirms our general statement \cite{YeZhangWang2022}
that an appropriate
choice of the regularization order generally depends upon
the spacetime background as well as upon the type of quantum fields,
and that the 4th-order  regularization may be necessary in some cases. 
In summary,  for the Maxwell field with the GF term  in the MD stage,
the total regularized vacuum stress tensor is zero,
and there is no trace anomaly,
and there is no need to introduce a ghost field
to cancel the GF vacuum stress tensor.
These features  are similar to those in the RD stage
and in the de Sitter space.

\end{document}